\documentstyle[12pt]{article}
\makeatletter
\@addtoreset{equation}{section}
\makeatother

\newcommand{\nc}{\newcommand}
\nc{\be}{\begin{equation}}
\nc{\ee}{\end{equation}}
\nc{\p}{\phi}
\nc{\m}{\mu}
\nc{\n}{\nu}
\begin{document}
\begin{titlepage}
\bigskip
\bigskip
\begin{center}
{\bf STRINGY ROBINSON-TRAUTMAN SOLUTIONS }\\
\bigskip
\bigskip
\bigskip
\bigskip
R. G{\" u}ven\\
\smallskip
Department of Mathematics,\\ Bo{\u g}azi{\c c}i University\\
Bebek, {\. I}stanbul 80815, Turkey\\
\bigskip
and\\
\bigskip

E. Y{\" o}r{\" u}k\\
Department of Physics,\\ Bo{\u g}azi{\c c}i University\\
Bebek, {\. I}stanbul 80815,Turkey\\
\bigskip
\bigskip
\bigskip
\bigskip
\bigskip
{\bf ABSTRACT}\\
\smallskip
\end{center}

   A class of solutions of the low energy string theory in four dimensions is 
studied. This class  admits a geodesic, shear-free null congruence which is 
non-twisting but in general diverging and the corresponding solutions in 
Einstein's theory form the Robinson-Trautman family together with a subset of 
the Kundt's class. The Robinson-Trautman conditions are found to be frame 
invariant in string theory. The Lorentz Chern-Simons three form of the stringy 
Robinson-Trautman solutions is shown to be always closed. The stringy 
generalizations of the vacuum Robinson-Trautman equation are obtained and 
three subclasses of solutions are identified. One of these subclasses exists, 
among all the dilatonic theories, only in Einstein's theory and in string 
theory. Several known solutions including the dilatonic black holes, the pp-
waves, the stringy C-metric and certain solutions which correspond to exact 
conformal field theories are shown to be particular members of the stringy 
Robinson-Trautman family. Some new solutions which are static or 
asymptotically flat and radiating are also presented. The radiating solutions 
have a positive Bondi mass. One of these radiating solutions has the property 
that it settles down smoothly to a black hole state at late retarded times.
\bigskip

\end{titlepage}

\section{Introduction}

\indent

     In general relativity Robinson-Trautman solutions \cite{1} have proven 
to be an interesting laboratory for addressing certain issues of black holes, gravitational
radiation and the asymptotic structure of space times. These solutions
are singled out by requiring that the spacetime admits a shear-free, non-twisting, geodesic null
congruence. In Einstein-Maxwell theory this requirement leads to a large class
of algebraically special solutions which belong to various Petrov types 
and one can identify the Reissner-Nordstr{\" o}m black holes, the charged 
C-metric as well as certain radiating solutions as particular members of 
the Robinson-Trautman family \cite{2}. The explicit forms of all the 
Robinson-Trautman solutions are, however, not known and the relevance of 
some of these solutions to the black hole formation \cite{3} as well as 
the structure of the Robinson-Trautman equations \cite{4} are 
still problems of current interest. 

    In this paper we wish to study the Robinson-Trautman solutions of string 
theory in four dimensions. We shall look for the solutions of the low 
energy string theory which admit, in the Einstein frame, a shear-free, 
non-twisting, geodesic null congruence. We shall derive the stringy 
generalization of the Robinson-Trautman equation and examine the various 
subclasses of solutions. One of these subclasses has the feature that it 
distinguishes, among all the dilatonic theories, Einstein`s theory and 
the string theory. This subclass exists only in these two theories. The 
whole family of stringy Robinson-Trautman solutions, however, turns out 
to be interesting for several other reasons. For example, we shall show 
that Robinson-Trautman conditions are frame invariant in string 
theory. Hence if a solution is of the Robinson-Trautman type in the 
Einstein frame, then it will also be a Robinson-Trautman solution in the 
string frame. It is also worth noting that these solutions do not admit 
in general any Killing vectors. Due to this property, the whole family 
cannot be generated from the vacuum solutions of Einstein`s equations by 
Ehlers-Harrison type transformations \cite{5}. Another attractive feature 
concerns the Lorentz Chern-Simons three-form. We shall prove that the 
Lorentz Chern-Simons three-form of the Robinson-Trautman family is always 
closed. This property is relevant to the higher order corrections and we 
shall see that the stringy Robinson-Trautman family contains, in fact, 
several exact solutions or the leading order representations of exact 
conformal field theories. One of the exact solutions which is of the 
Robinson-Trautman type is that of an electrically charged, extreme black 
hole \cite{6}, \cite{7}. Another such solution can be identified by noting 
that, in contrast to Einstein-Maxwell theory and because of the presence 
of the dilaton, it is possible to switch off the divergence of the 
geodesic null congruence within the stringy Robinson-Trautman family. By 
specializing to this case and choosing the spacetime to be conformally 
flat, one can arrive at the special plane wave solution which is 
interpretable as a WZW model \cite{8} and plane waves are also known to 
be exact solutions \cite{9}. These two examples show that the 
Robinson-Trautman family has a non-empty intersection with the chiral 
null models \cite{10}. We shall recover several other known solutions as 
particular members of the Robinson-Trautman family. It will be seen that 
the family contains the set of all charged dilatonic black holes \cite{6}, 
the pp-waves \cite{11}, the stringy C-metric \cite{12}, the static, 
spherically symmetric solutions of \cite{13} as well as the solutions of 
\cite{14} which correspond to exact conformal field theories. We shall also 
present explicitly some new solutions which are static or asymptotically flat 
and radiating. The radiating solutions that we shall study have a positive 
Bondi mass. One of these radiating solutions tends smoothly to the stringy 
black holes at late retarded times.

\section{Robinson-Trautman Form of the Fields}

\indent

Most of our discussions are based on the action
\be\label{2.1}
S=\int\ 
d^{4} x 
\sqrt{-g}\left[R+2\nabla_{\m}\p\nabla^{\m}\p
-\tilde{\kappa}^{2}e^{-2a\p}F_{\m\n}F^{\m\n} 
\right], 
\ee
where \( R\) is the Ricci scalar for the metric \( g_{\m\n}\), the real scalar 
field \(\p\) is the dilaton, 
\( F_{\m\n}=\partial_{\m}A_{\n}-\partial_{\n}A_{\m}\)  is the spin-1 field 
strength,  \(\tilde{\kappa}\) is the coupling constant, \( g=det(g_{\m\n})\) 
and \(a\) is a real parameter. When \(0 \leq a \leq 1\), this action describes 
the dilatonic gravity theories which interpolate between the standard 
Einstein-Maxwell theory \((a=0,\phi=const.\)) and the low-energy string theory 
\((a=1)\). The field equations that follow from (2.1) can be written as
\be\label{2.2}
d\star\left(e^{-2a\p}F\right)=0,
\ee
\be\label{2.3}                   
d\star d\p-a\tilde{\kappa}^{2}e^{-2a\p}F\wedge \star 
F=0,
\ee
\be\label{2.4}
R_{\m\n}=-2\nabla_{\m}\p\nabla_{\n}\p
+2\tilde{\kappa}^{2}e^{-2a\p}\left(F_{\m\lambda}F_{\n}^{\lambda}
-\frac{1}{4}g_{\m\n}F_{\kappa\lambda}F^{\kappa\lambda}\right),
\ee
where \(F=\frac{1}{2}F_{\m\n}   dx^{\m} \wedge dx^{\n}\) is the Maxwell 
two-form, \(R_{\m\n}\) is the Ricci tensor and \( \star \) denotes the 
Hodge 
dual. It can be checked that these equations are invariant under 
the duality transformations
\be\label{2.5}
F \to e^{-2a\p} \star F,~~~~ \p \to  -\p .
\ee
We shall be primarily concerned with the \(a=1\) case of (2.1). In 
identifying this case as the low-energy string action one assumes that 
the axion field is set equal to zero. The axion field strength, however, 
involves the spin-1 Chern-Simons three-form and for a proper 
identification, the field equations must be complemented with
\be\label{2.6}
F \wedge F=0.
\ee
Since the gravitational part of (2.1) has the standard Einstein-Hilbert 
form, \(g_{\m\n}\) is the Einstein metric and the \(a=1\) specializations 
of (2.2)-(2.4) together with (2.6) are the string field equations in the 
Einstein frame. To pass to the string frame, one must introduce at \(a=1\)
\be\label{2.7}
g^{s}_{\m\n}=e^{2a\p}g_{\m\n},
\ee
and transform (2.1) so that the string metric \(g^{s}_{\m\n}\) is the 
gravitational field variable.

Our first goal is to characterize the Robinson-Trautman solutions of the 
string field equations in the Einstein frame. For this purpose it will be 
convenient to employ the Newman-Penrose (NP) formalism \cite{15} and work 
with a null tetrad \((l_{\m},n_{\m},m_{\m},\bar {m}_{\m})\). The null 
tetrad determines the metric as 
\be\label{2.8}
g_{\m\n}=2l_{(\m}n_{\n)}-2m_{(\m} \bar{m}_{\n)}.
\ee
The NP form of the Cartan`s equations of structure and the decomposition 
of the spacetime curvature are described in the Appendix. The spin-1 
field, on the other hand, is represented by three complex scalars 
\(\Phi_{0}, \Phi_{1}, \Phi_{2}\) and these may be defined by
\be\label{2.9}
F+i\star F=-2\Phi_{1}(l \wedge n-m \wedge \bar{m})-2\Phi_{0}n \wedge 
\bar{m}+2\Phi_{2}l \wedge m,
\ee
where \(l=l_{\m}dx^{\m}\), \(n=n_{\m}dx^{\m}\), \(m=m_{\m}dx^{\m}\) are 
the null basis one forms; \(l\) and \(n\) are real, \(m\) is complex.
Throughout  the paper an overbar denotes complex conjugation. We shall choose 
\(l^{\m}\) to be tangent to a geodesic, shear-free, non-twisting null 
congruence. In terms of the NP spin coefficients, this means
\be\label{2.10}
\kappa=\sigma=0 ,~~~~ \rho=\bar{\rho} .
\ee
We shall also assume that \(l^{\m}\) is a null eigenvector 
of \(F_{\m\n}\):
\be\label{2.11}
\Phi_{0}=0 ,
\ee
and impose on the full energy momentum tensor the conditions
\be\label{2.12}
\Phi_{01}=\Phi_{02}=0 .
\ee
If one were to impose (2.10) and (2.11) in Einstein-Maxwell theory, the 
spacetime would be algebraically special:
\be\label{2.13}
\Psi_{0}=\Psi_{1}=0 ,
\ee
where \(\Psi_{0}, \Psi_{1}\) are Weyl scalars (see Appendix) and (2.12) would 
follow from (2.11). What one is really dealing with would then be a particular 
case of the Goldberg-Sachs theorem \cite{2}. Because the dilaton also 
contributes to the energy-momentum tensor, this is no longer the case in 
string theory. We shall require the stringy Robinson-Trautman family to 
share the algebraic character of its Einstein counterpart and impose  
(2.13) as an additional condition on the spacetime curvature. Hence the 
solutions which obey (2.10)-(2.13) will constitute the stringy 
Robinson-Trautman family.

The null vector \(l^{\m}\) is now an eigenvector of both \(F_{\m\n}\) and 
the Weyl tensor and the tetrad gauge freedom is partially fixed. There is 
still the freedom of performing null rotations which preserve the 
direction of \(l^{\m}\) and these involve four real parameters. Using 
such a null rotation the complex connection one-form  \(\Gamma_{1}\) defined, 
in the Appendix, can be reduced to 
\be\label{2.14}
\Gamma_{1}=\rho m ,
\ee
and as a consequence of (2.12) and (2.13) one finds
\be\label{2.15}
\Gamma_{1} \wedge d \Gamma_{1}=0 .
\ee
The Frobenius theorem then allows one to introduce two arbitrary 
functions \(P_{0}\) and \(z\) such that
\be\label{2.16}
\Gamma_{1}=(\rho/P_{0})d \bar{z}
\ee
and \(P_{0}\) can always be choosen to be real. Then, according to (2.14) 
and (2.16), the complex leg of the null tetrad must have the form
\be\label{2.17}
m=\frac{1}{P_{0}}d \bar{z} .
\ee
Since (2.10) is assumed to hold, \(l\) also satisfies
\be\label{2.18}
l \wedge dl=0
\ee
and the Frobenius theorem can be utilized once again together with a null 
rotation which preserves (2.17) to set
\be\label{2.19}
l=du ,
\ee
where \(u\) is an arbitrary real function. In order to determine the form 
of the final leg \(n\), let us choose as our coordinate system 
\((u,r,z,\bar{z})\), where \(r\) is an affine parameter along the null 
geodesic tangent to \(l^{\m}\). In such a coordinate system the most general 
\(n\) can be written as
\be\label{2.20}
n=dr+H_{0}du+Wd \bar{z}+\bar{W}dz ,
\ee
where \(H_{0}\) is a real function and \(W\) is complex. Computing now 
all the NP spin coefficients for (2.17), (2.19) and (2.20) gives
\be\label{2.21}
W'=0   ,~~~~W_{z}=\bar{W}_{\bar{z}} .
\ee
Here and in the sequel we use primes to denote differentiation with 
respect to \(r\) whereas the other partial derivatives are denoted by 
subscripts: \(W'=\partial W/\partial r , W_{z}=\partial W/\partial z , 
W_{u}=\partial W/\partial u\). It follows 
from the conditions (2.21) that one can reduce (2.20) to the form 
\be\label{2.22}
n=dr+Hdu ,
\ee
by a coordinate transformation: \(r \to  r+f(u,z,\bar{z})\) without altering 
the forms of \(l\) and \(m\). When the last two terms of (2.20) are 
gauged away by setting  \(f_{z}=\bar{W}\), the coefficient of \(du\) 
transforms into a new function \(H\) and \(r\) is now the new coordinate. 
Hence there exists a coordinate system \((u,r,z,\bar{z})\) where one is 
left only with two real functions \(P_{0}\) and \(H\). In this coordinate 
system (2.10)-(2.13) are all satisfied by an arbitrary \(H\) but \(P_{0}\)
must obey
\be\label{2.23}
(lnP_{0})_{z}'=0 ,
\ee
in order to fulfill \(\Psi_{1}=0\). Thus the final outcome of the 
conditions (2.10)-(2.13) can be written as
\begin{eqnarray}\label{2.24}
l&=&du ,\nonumber\\
n&=&dr+Hdu ,\nonumber\\
m&=&\frac{R}{P}d \bar{z} ,
\end{eqnarray}
where \(H=H(u,r,z,\bar{z})\), \(R=R(u,r)\), \(P=P(u,z,\bar{z})\) are real 
functions. The stringy Robinson-Trautman line element must therefore be 
of the form
\be\label{2.25}
ds^{2}=2dudr+2Hdu^{2}-2R^{2}\frac{dzd \bar{z}}{P^{2}} .
\ee

\indent

The conditions (2.11) and (2.12) have also nontrivial implications on the 
dilaton through the field equation (2.4). The vanishing of the 
\(\Phi_{01}\) component of the Ricci tensor requires \(\p'\p_{z}=0\) and 
\(\Phi_{02}=0\) implies \(|\p_{z}|^{2}=0\). Hence the allowable dilaton 
fields are of the form 
\be\label{2.26}
\p=\p(u,r)
\ee
and we now have the most general forms of \(g_{\m\n}\), \(\p\) and 
\(F_{\m\n}\) that belong to the Robinson-Trautman family. Notice that 
this family is defined for all dilatonic theories.

Let us next impose (2.6). For an arbitrary Maxwell field and in terms of 
the NP variables, (2.6) amounts to
\be\label{2.27}
\Phi_{1}^{2}-\Phi_{0}\Phi_{2}=\bar{\Phi}_{1}^{2}-\bar{\Phi}_{0}\bar{\Phi}_{2} .
\ee
In our case \(\Phi_{0}=0\) and therefore, \(\Phi_{1}=\pm\bar{\Phi}_{1}\). 
Since the electric case: \(\Phi_{1}=\bar{\Phi}_{1}\) and the magnetic 
case: \(\Phi_{1}=-\bar{\Phi}_{1}\) can be mapped to one another by a 
duality transformation (2.5), we shall choose to work only with the 
magnetic solutions:

\be\label{2.28}
\Phi_{1}=-\bar{\Phi}_{1} .
\ee

The problem is now to determine two real functions \(R, \p\) of two 
variables, a real function \(P\) of three variables together with a real 
\(H\), a purely imaginary \(\Phi_{1}\) and a complex \(\Phi_{2}\) where 
the last three functions can depend on all the coordinates. Before 
embarking into the differential equations which govern these functions 
three points are worth noticing:

First, if one compares the above results with those of the 
Einstein-Maxwell theory, one finds as the new feature of the stringy 
Robinson-Trautman metric, the presence of a general warp factor 
\(R(u,r)\). As we shall see below this warp factor is coupled to the 
dilaton. Whenever \(\p=const.\), \(R''=0\) and without any loss of 
generality one can take  \(R=r\) in Einstein-Maxwell theory. According 
to (2.3) it is, of course, not permissible to set \(\p=const.\) unless \(a=0\) 
or \(F \wedge \star F=0\). In Einstein's theory the metric (2.25) was 
previously encountered in perfect fluid solutions \cite{16}.

Secondly, suppose one chooses to work with the string metric 
\(g^{s}_{\m\n}\) rather than the Einstein metric. The starting point of 
the foregoing analysis will then be the null tetrad of \(g^{s}_{\m\n}\). 
Because of (2.7), this tetrad can be related to (2.24) by the conformal 
transformation
\be\label{2.29}
l^{s}=e^{a\p}l ,~~ n^{s}=e^{a\p}n ,~~ m^{s}=e^{a\p}m .
\ee
It can be checked by computing the connection and the curvature of the 
new tetrad that, as long as \(\p=\p(u,r)\), the conditions (2.10)-(2.13) 
will remain to hold in the string frame. Hence if a solution is of the 
Robinson-Trautman type in the Einstein frame, it will also be a 
Robinson-Trautman solution in the string frame.

Finally, let us consider the Lorentz Chern-Simons three-form in the 
Einstein frame. The Lorentz Chern-Simons three-form \(\omega^{0}_{3L}\) 
is known to satisfy
\be\label{2.30}
d \omega^{0}_{3L}=\Omega_{ab} \wedge \Omega^{ab} ,
\ee
where \(\Omega_{ab}\) are the curvature two-forms. In terms of the NP 
variables one finds that
\be\label{2.31}
\star d 
\omega^{0}_{3L}=4i\left[3(\bar{\Psi}_{2}^{2}-\Psi_{2}^{2})
+(\bar{\Psi}_{0}\bar{\Psi}_{4}
-\Psi_{0}\Psi_{4})+4(\Psi_{1}\Psi_{3}
-\bar{\Psi}_{1}\bar{\Psi}_{3})\right] .
\ee
When one specializes to the Robinson-Trautman metrics, 
\(\Psi_{0}=\Psi_{1}=0\) and moreover, it turns out that 
\(\Psi_{2}=\bar{\Psi}_{2}\) (see Appendix). Hence for any metric having 
the form (2.25)
\be\label{2.32}
d \omega^{0}_{3L}=0
\ee
and consequently, the axion field which was taken to be zero in the 
leading order approximation, can be maintained to be zero even after 
including \(\omega^{0}_{3L}\) as a higher order correction. Notice that 
(2.25) is more general than the metrics which are known to enjoy this 
property \cite{17}.

\section{ The Field Equations}   

\indent

We are now in a position to consider the field equations which govern the 
Robinson-Trautman form of the fields. Among these Maxwell equations (2.2) 
are the simplest:
\be\label{3.1}
(R^{2}\Phi_{1})'=0 ,
\ee
\be\label{3.2}
(R\Phi_{2})'=0 ,
\ee
\be\label{3.3}
\Phi_{1z}+aR\p'(\bar{\Phi}_{2}/P)=0 ,
\ee
\be\label{3.4}
(R^{2}\Phi_{1}/P^{2})_{u}+R(\Phi_{2}/P)_{z}=0 .
\ee
The first two equations can be readily integrated to give
\be\label{3.5}
\Phi_{1} =\frac{c}{R^{2}},~~ \Phi_{2}=\frac{h}{R},
\ee
where \(c=c(u,z,\bar{z})\), \(h=h(u,z,\bar{z})\) and because (2.28) 
holds, \(c=-\bar{c}\). Hence \(\Phi_{1}\) and \(\Phi_{2}\) depend on the 
coordinate \(r\) only through the function \(R(u,r)\). When (3.5) is 
taken into account, (3.3) and (3.4) become
\be\label{3.6}
Pc_{z}=-a\bar{h}R^{2}\p',
\ee
\be\label{3.7}
(c/P^{2})_{u}+(h/P)_{z}=0.
\ee
The \(\Phi_{00}\) component of the Einstein equation (2.4) has also a 
very simple form:
\be\label{3.8}
R''+(\p')^{2}R=0,
\ee
and the other non-trivial components of (2.4) can be summarized as
\be\label{3.9}
6\Lambda=(H\p'-\p_{u})\p',
\ee
\be\label{3.10}
\Phi_{11}+3\Lambda=2\tilde{\kappa}^{2}e^{-2a\p}|c|^{2}/R^{4},
\ee
\be\label{3.11}
\Phi_{12}=2\tilde{\kappa}^{2}e^{-2a\p}\bar{h}c/R^{3},
\ee
\be\label{3.12}
\Phi_{22}=(H\p'-\p_{u})^{2}+2\tilde{\kappa}^{2}e^{-2a\p}|h|^{2}/R^{2},
\ee
where \(\Lambda,\Phi_{11},\Phi_{12},\Phi_{22}\) are constructed from the 
metric (2.25) and given in the Appendix. Finally, we have the dilaton 
field equation (2.3) and this reduces to

\be\label{3.13}
P^{2}[(R^{2}/P^{2})\p']_{u}+[R^{2}(\p_{u}-2H\p')]'
=-4a\tilde{\kappa}^{2}e^{-2a\p}c^{2}/R^{2}.
\ee
The problem is now to determine the six functions \((R,\p),(P,c,h)\) and 
\(H\) as the solutions to these equations. In this process it is useful 
to note two distinct symmetries of the problem. One of these is the 
freedom of rescaling the functions:
\be\label{3.14}
R \to w(u)R,~~P \to w(u)P,~~c \to w^{2}c,~~h \to w(u)h
\ee
by an arbitrary \(u\)-dependent function \(w(u)\). It is straightforward 
to check that the metric (2.25), the Maxwell field (3.5) and the field 
equations (3.6)-(3.13) are all form-invariant under (3.14). The second 
symmetry concerns the coordinate gauge freedom of the metric (2.25) and 
the corresponding tetrad rotations of (2.24). Of particular interest is 
the coordinate transformation:

\be\label{3.15}
u \to \tilde{u}=\int\ f^{-1}(u)du,
\ee
\be\label{3.116}
r \to \tilde{r}=f(u)r+g(u),
\ee
where \(f(u)\) and \(g(u)\) are arbitrary functions. Under such a 
transformation the metric function \(H\) transforms to

\be\label{3.17}
H \to \tilde{H}=f^{2}H-(\tilde{r}-g)f_{u}-fg_{u},
\ee
but the metric (2.25) preserves its form. This coordinate transformation 
induces on the tetrad (2.24) the null rotation:

\be\label{3.18}
l \to \tilde{l}=f^{-1}l,~~ n \to \tilde{n}=fn,~~ \tilde{m}=m,
\ee
which preserves \(m\) as well as the directions of \(l\) and \(n\). The 
rescalings (3.14) and the transformations (3.15)-(3.17) turn out to be  
valuable tools in handling the arbitrary functions that arise  through 
integrations. It may also be  of interest to note that the coupling constant 
\(\tilde{\kappa}\) can always be set equal to one by adding an appropriate 
constant to \(\p\). This is, of course, manifest in (2.1).

\section{The Case \(\p' \neq 0, h \neq 0\)}
    
\indent

Consider now (3.6). Provided \(\p' \neq 0\), this equation determines 
\(h(u,z,\bar{z})\) in terms of the other functions and thereby reduces 
the unknowns by one. Since the left hand side has no \(r\)-dependence, 
it also requires, unless \(h=0\), that

\be\label{4.1}
2aR^{2}\p'=-U(u),
\ee
where \(U(u)\) is an arbitrary function. With the above choice of the 
factors in (4.1) one has \((e^{-2a\p})'=Ue^{-2a\p}/R^{2}\) and the 
dilaton equation (3.13) can be integrated with respect to \(r\). The 
result is an expression for \(H+r(lnP)_{u}\). Forming \(\Phi_{12}\) from 
this expression and comparing with the right hand side of (3.11), where now 
\(\bar{h}=2Pc_{z}/U\), shows that (3.11) is satisfied if and only if

\be\label{4.2}
(a^{2}-1)cc_{z}=0.
\ee 
Recall that the parameter \(a\) labels the different dilatonic theories 
and \(c_{z} \neq 0\) by assumption. Hence the Robinson-Trautman solutions 
with \(\p' \neq 0, h \neq 0\) exist, among all the dilatonic theories, 
only in string theory: \(a=1\). A similar set of solutions also exists in 
the Einstein-Maxwell theory where \(a=c_{z}=0\) and  \(\p=const.\) but not in 
any other theory with \(a \neq 1\). Having noted this interesting 
selection within the dilatonic theories, from now on we shall assume that

\be\label{4.3}
a=1.
\ee
The other subfamilies of solutions having \(h=0\) or \(\p'=0\) which we 
shall later study in fact allow generalizations to arbitrary values of 
\(a\). However, we shall not be concerned with these generalizations and 
concentrate on the solutions of the string theory. 

After substituting \(\p' = -U/2R^{2}\) and integrating once with respect to 
\(r\), (3.8) can be written as

\be\label{4.4}
[(R^{2})']^{2}=4s(u)R^{2}+U^{2}(u),
\ee
where \(s(u)\) arises as a function of integration. Depending on whether 
\(s(u)\) can be taken to be zero or not, two subcases need to be distinguished.
\newpage
{\bf The   Subcase}    \(s(u)~ \neq ~0\) :

\indent

When \(s(u) \neq 0\) both \(\p\) and \(R\) can be completely 
determined, all \(r\) integrations can be completed and the problem 
reduces to the solution of two coupled partial differential equations. 
Starting from (4.4) and utilizing the gauge freedom (3.14)-(3.18), one 
finds that without any loss of generality one can set

\be\label{4.5}
e^{-2\p}=b-\frac{k}{r},~~~ R=re^{-\p},
\ee
where \(b\) and \(k\) are constants. After introducing two real functions 
\(Q(u,z,\bar{z})\) and \(M(u,z,\bar{z})\) which satisfy

\be\label{4.6}
Q^{2}=kM
\ee
where \(Q=-2i\tilde{\kappa}c\) and the operator

\be\label{4.7}
\Delta=2P^{2}\partial_{z} \partial_{\bar{z}},
\ee
which is essentially the Laplacian on the \(u=const.\), \(r=const.\) 
hypersurfaces, one obtains

\be\label{4.8}
\Phi_{1}=\frac{i}{2\tilde{\kappa}} \frac{Q}{R^{2}},~~~~
\Phi_{2}=-\frac{i}{\tilde{\kappa}} \frac{PQ_{\bar{z}}}{kR},
\ee
\be\label{4.9}
H=\frac{1}{2b}[\Delta lnP+k(lnP)_{u}]-\frac{M}{r}-r(lnP)_{u}.
\ee
The solutions belonging to this category therefore involve two parameters 
\(b, k\) and two independent functions \(P,M\). It is possible to set \(b=1\) 
by choosing \(\p \to 0\) as \(r \to \infty\) and, of course, one can choose to 
work with \(Q\) in place of \(M\). The functions \(P\) and \(M\) are governed 
by

\be\label{4.10}
\Delta M+k[4M(lnP)_{u}-M_{u}]=\frac{P^{2}}{M} M_{z}M_{\bar{z}},
\ee
\be\label{4.11}
\Delta \Delta 
(lnP)+12b^{2}M(lnP)_{u}-4b^{2}M_{u}+k^{2}[2((lnP)_{u})^{2}-(lnP)_{uu}]
=\frac{2b^{2}P^{2}}{kM} M_{z}M_{\bar{z}}
\ee
and it can be checked that all field equations are taken into account. 
Equations (4.10) and (4.11) are the stringy ge\-ne\-ral\-iza\-tions of the va\-cuum 
Ro\-bin\-son-Traut\-man equation. When \(k=0\), \(Q=0\) and it can also be 
deduced that \(M_{z}=0\). In this limit (4.10) is trivially satisfied and (4.11) 
reduces to the standard form of the vacuum Robinson-Trautman equation \cite{2}  
by setting \(b=1\). Notice that because \(Q^{2}=kM\), 
one can replace (4.10) with

\be\label{4.12}
\Delta Q=kP^{2}(Q/P^{2})_{u}.
\ee

In general, the solutions under consideration are algebraically 
special, admit no Killing vectors and can belong to various Petrov types. 
Suppose we now concentrate on the solutions obeying \(P_{u}=0\). Then 
(4.12) is a heat equation on a two dimensional surface having the metric 
\(ds^{2}=2dzd\bar{z}/P^{2}\) and \(k\) plays the role of a diffusion 
constant. The Gaussian curvature of this two-surface is

\be\label{4.13}
K=\Delta lnP.
\ee
Let us specialize further to the spaces of constant curvature and 
normalize \(K\) to \(K=0,\pm 1\). Under these assumptions 

\be\label{4.14}
P=1+\frac{1}{2}Kz \bar{z}
\ee
and the other metric function can be written as

\be\label{4.15}
2H=K-\frac{2M}{r},
\ee
where we have taken \(b=1\). The simplest way to satisfy (4.10) and 
(4.11) is then to set \(M=const.\) and clearly, if \(M > 0\), \(k \geq 
0\). We have now obtained three different Robinson-Trautman solutions
depending on the value of \(K\) and they are all of Petrov type D.
The significant solution among these has \(K=1\) and describes the dilatonic 
black holes \cite{6} if \(Q^{2} \leq 2M^{2}\).

Another solution which belongs to this subcase and which is also of Petrov 
type D is the dilatonic C-metric \cite{12}. In this particular
example both \(P\) and \(Q\) depend solely on a function \(x=x(u,z,\bar{z})\)  
which is defined implicitly by 
\be\label{4.16}
G^{-1}(x) dx = \frac{1}{2}(dz+d \bar{z})+Adu,
\ee
where
\be\label{4.17}
G(x)=1-x^{2}-Ar_{+}x^{3}
\ee
and \(A,r_{+}\) are real parameters. After choosing  \(b=1\) and introducing
\be\label{4.18}
\tilde{F}(x)=1+kAx,
\ee
one has
\be\label{4.19}
P^{2}=2 \tilde{F}(x)/G(x),~~  Q=-q \tilde{\kappa}\tilde{F}(x),
\ee
where \(q\) is the fourth parameter appearing in the solution. The metric 
function \(H\) is now expressible as
\be\label{4.20}
H=\frac{r^{2}A^{2}}{2 \tilde{F}(x)}\left[e^{-2\p}G(x)-G \left(xe^{-2\p}
-\frac{1}{Ar}\right) \right].
\ee

{\bf The  Subcase}   \(s(u)~=~0\)  :

\indent

Starting from (4.4) one can determine the \(r\)-dependence of all the fields
completely also when \(s(u)=0\) but now a different structure emerges. This 
time the gauge freedom (3.14)-(3.17) allows one to set
\be\label{4.21}
\p=\p_{0}(u)-\frac{\epsilon}{2}lnr,~~ R^{2}=r,
\ee
where \(\p_{0}(u)\) is arbitrary, \(\epsilon=\pm1\) and the coordinate \(r\) 
is restricted to the range \(r > 0\). Introducing again two functions 
\(Q(u,z,\bar{z})\) and \(M(u,z,\bar{z})\) one obtains
\be\label{4.22}
\Phi_{1}=\frac{i}{2\tilde{\kappa}} \frac{Q}{R^{2}},~~ 
\Phi_{2}=\frac{-i}{\tilde{\kappa}} \frac{Q_{\bar{z}}}{R}
\ee
and
\be\label{4.23}
H=M+Q^{2}e^{-2\p_{0}}r^{\epsilon}-r[(lnP)_{u}+ \epsilon \dot{\p}_{0}],
\ee
where \(\dot{\p}_{0}=d \p_{0}/du\). The field equations no longer impose a 
relationship between \(Q\) and \(M\) and reduce to the following:
\be\label{4.24}
\Delta Q-\epsilon P^{2}(Q/P^{2})_{u}=0,
\ee
\be\label{4.25}
\Delta M+P^{2}(M/P^{2})_{u}-2\epsilon \dot{\p}_{0}M=0,
\ee
\be\label{4.26}
\Delta lnP-(lnP)_{u}+\epsilon \dot{\p}_{0}=(1+\epsilon)Q^{2}e^{-2\p_{0}},
\ee
where the operator \(\Delta\) is again defined by (4.7).

When all fields are assumed to have no u-dependence, (4.24)-(4.26) take a 
particularly simple form and become equations  on the two-dimensional 
spacelike surface whose Gaussian curvature is given by (4.13). According to 
(4.26), the Gaussian curvature is now equal to
\be\label{4.27}
K=(1+\epsilon)Q^{2}e^{-2\p_{0}}
\ee
and the surface in question is locally flat if \(\epsilon=-1\). 
For both values of \(\epsilon\), \(M\) and \(Q\) are two independent harmonic 
functions on this two-surface and moreover, \(K \geq 0\). Specializing to the 
spaces of constant curvature, if \(K=0\), either \(\epsilon=-1\) or 
\(\epsilon=1\), \(Q=0\). The two-sphere, \(K=1\), is allowed only if
\(\epsilon=1\) in which case \(2Q^{2}=e^{2\p_{0}}\) and \(M\) must also be 
constant. Note that whenever \(s(u)=0\), the string frame line element is

\be\label{4.28}
ds_{s}^{2}=e^{2\p_{0}} \left[ 2r^{- \epsilon}(dudr+Hdu^{2})
-2 \frac{r^{1- \epsilon}}{P^{2}}dzd \bar{z} \right]
\ee
and if the fields are independent of \(u\) and \(\epsilon=1\), the string
metric is a direct product of two two-dimensional metrics. Choosing \(K=1\) 
gives in particular
\be\label{4.29}
ds_{s}^{2}=e^{2\p_{0}} \left[\frac{2}{r}dudr
+ \left(1+ \frac{2M}{r} \right)du^{2}
-d \Omega_{2}^{2} \right],
\ee
where  \(d \Omega_{2}^{2}\) is the line element of the unit two-sphere. The 
sign of the constant \(M\) now determines the sign of the Gaussian curvature 
of the \((u,r)\)-subspace. Choosing \(M=0\) produces the throat solution of 
\cite{14} and by letting \(2M=-e^{-\p_{0}}\) one can check that (4.29) reduces 
to the black hole plus infinite throat solution of \cite{14}. The third, 
asymptotically flat region plus infinite throat solution of \cite{14}, on the 
other hand, can be regained from the previous subcase \(s(u) \neq 0\). Each of 
these three solutions is known to correspond to an exact conformal field 
theory \cite{14}.

\section{The Case  \(h=0\), \(\p' \neq 0\)}

\indent

When \(h=0\) the two principal null directions of the Maxwell tensor coincide: 
\(\Phi_{2}=0\). Clearly, in this case \(c_{z}=0\) and (4.1) cannot be deduced 
from (3.6). From (3.7) and (3.11) it follows that
\be\label{5.1}
(lnP)_{uz}=0,~~ H_{z}'=0.
\ee
Differentiating the dilaton equation (3.13) with respect to \(z\) then gives 
the condition
\be\label{5.2}
H_{z}(R^{2} \p')'=0,
\ee
which can be satisfied in two ways. If \(H_{z} \neq 0\), one sees that (4.1) 
must hold. Hence the solutions with \(H_{z} \neq 0\) are simply the 
\(\Phi_{2}=0,~Q=Q(u)\) specializations of the solutions discussed in 
Section IV. Note that for these special solutions (4.2) holds for all values 
of the parameter \(a\).

The second way to fulfill (5.2) is, of course, to set \(H_{z}=0\). Then (4.1) 
need not hold and one is dealing with a subset which contains new solutions. 
All of these solutions are of Petrov type D and can be represented by the 
fields having the form
\be\label{5.3}
H=H(u,r),~~ P=P(z,\bar{z}),~~ \Phi_{1}=R^{-2}(u,r),
\ee
modulo the gauge freedom (3.14)-(3.17). The two-dimensional 
\(u=const.\), \(r=const.\) sections of these spacetimes are again spaces of 
constant Gaussian curvature:
\be\label{5.4}
\Delta lnP=K
\ee
and the functions \( H(u,r),\p(u,r),R(u,r) \) are governed by the field 
equations
\be\label{5.5}
R''+( \p ')^{2}R=0,
\ee
\be\label{5.6}
(R^{2} \p ')_{u}+ \left[ R^{2}( \p_{u}-2H \p ') \right]'
=-4 \tilde{\kappa}^{2}e^{-2 \p }/R^{2},
\ee
\be\label{5.7}
K+(R^{2})_{u}'-\left[ H(R^{2})' \right]'=
-4\tilde{\kappa}e^{-2 \p }/R^{2},
\ee
\be\label{5.8}
-K+ \left( R^{2}H \right)''=2R^{2} \p ' \p_{u}+ \left( R^{2} \right)_{u}'
+2R^{2}(lnR)_{u}'+ \left( R^{2} \right)'(lnR)_{u},
\ee
\be\label{5.9}
H_{u}R'+2HR_{u}'-H'R_{u}-R_{uu}=( \p_{u}-2H \p ')R \p_{u}.
\ee
For the special case \(K=1,H_{u}=R_{u}=\p_{u}=0\), the solutions of these 
equations were obtained in \cite{13} and it can be verified that similar sets 
of \(u\)-independent solutions also exist for \(K=0\) and \(K=-1\).

\section{The Case \(\p'=0\)}

\indent

When the dilaton depends only on \(u\) the dilaton equation (3.13) together 
with (3.8) can be used to infer
\be\label{6.1}
\Phi_{1}=0,~~~~R=R(u)
\ee
and because (3.14) is a symmetry of the problem, one can simply set
\be\label{6.2}
R=1.
\ee
This shows how the divergence of the null geodesic \(l^{\m}\) congruence is 
regulated in string theory by the dilaton field. When \(\p'=0\), it follows 
from (6.2) that \(\rho=0\) and one is dealing with non-diverging solutions 
which are analogs of Kundt's class in general relativity \cite{2}. This is to 
be contrasted with Einstein-Maxwell theory where a passage from the 
Robinson-Trautman family to Kundt's class is not possible.

For the case \(\p'=0\), all the field equations except (3.12) amount to
\be\label{6.3}
\Delta lnP=0,
\ee
\be\label{6.4}
H=V(u,z,\bar{z})-r(lnP)_{u},
\ee
\be\label{6.5}
\Phi_{2}=P \bar{\Sigma}  (u,\bar{z})
\ee
The general solution of (6.3) is well-known:
\be\label{6.6}
lnP=f(u,z)+ \bar{f}(u, \bar{z})
\ee
where \(f(u,z)\) is an arbitrary complex function which is analytic in \(z\). 
The function \(\Sigma(u,z)\) is also analytic in \(z\) but otherwise 
arbitrary. Since the dilaton field is not constrained by the field equations, 
the solutions involve three arbitrary functions \(\p(u)\), \(f(u,z)\) and 
\(\Sigma(u,z)\). The remaining field equation (3.12) becomes
\be\label{6.7}
\Delta V= 2 \p_{u}^{2}+4 \tilde{\kappa}^{2} e^{-2\p}P^{2}|\Sigma|^{2}
+ \left[\left(lnP^{2} \right)_{u} \right]^{2}- \left(lnP^{2} \right)_{uu},
\ee
where the functions \(\p(u)\), \(f(u,z)\) and \(\Sigma(u,z)\) act as source 
terms. Hence the problem is now reduced to the solution of one non-trivial 
differential equation (6.7).

It can be checked that such solutions belong to one of the Petrov types 
III, N or O:
\be\label{6.8}
\Psi_{2}=0,
\ee
\be\label{6.9}
\Psi_{3}=P(lnP)_{u \bar{z}},
\ee
\be\label{6.10}
\Psi_{4}= \left(P^{2}V_{\bar{z}} \right)_{\bar{z}}
-r \left[P^{2}(lnP)_{u \bar{z}} \right]_{\bar{z}}
\ee
and the curvature scalar vanishes: \(\Lambda=0\). When \(\Psi_{3}=0\), one can 
set \(P=1\) by a coordinate transformation and a redefinition of the metric 
function \(V(u,z,\bar{z})\). In this special case the solutions are of Petrov 
type N and describe pp-waves. The pp-waves are known to be exact solutions 
of d=4 string theory when \(\p(u)\) is choosen appropriately \cite{11}, 
\cite{9}. If one further specializes to the conformally flat case 
\(V_{\bar{z} \bar{z}}=0\), the solution can be brought to the form
\be\label{6.11}
ds^{2}=2du \left[dr+\left(\p_{u}^{2}
+2 \tilde{\kappa}^{2}e^{-2\p}|\Phi_{2}|^{2} \right)z \bar{z}du \right]
-2dzd \bar{z},
\ee
\be\label{6.12}
\p = \p (u),~~ \Phi_{2}= \Phi_{2}(u),
\ee
where \(\p(u)\) and \(\Phi_{2}(u)\) are arbitrary. When \(\p\) and 
\(\Phi_{2}\) are constants such that 
\(8 \tilde{\kappa}^{2}e^{-2\p}|\Phi_{2}|^{2}=1\), this particular case can be 
interpreted as a WZW model which is based on the six dimensional Heisenberg 
group \cite{18}, \cite{8} and reduced to four spacetime dimensions.

\section{New Radiating Solutions}

\indent

Having seen all possible subfamilies of the stringy Robinson-Trautman family, 
we now present some new radiating solutions. The solutions that we shall 
consider belong to the subcase \(s(u) \neq 0\) of Section IV. Before deriving 
these solutions it will be instructive to go to a gauge where the dilaton 
picks up a \(u\)-dependence:
\be\label{7.1}
e^{-2\p}=b- \frac{D(u)}{r}~,~~R=re^{-\p}.
\ee
This is accomplished by a coordinate transformation (3.15)-(3.17) which has a  
simple effect on (4.6)-(4.9). All expressions in (4.6)-(4.9) remain valid 
except that now
\be\label{7.2}
Q^{2}(u,z, \bar{z})~=~D(u)M(u,z, \bar{z}),
\ee
\be\label{7.3}
H= \frac{1}{2b} \left[ \Delta lnP+D(lnP)_{u}- \dot{D} \right]- \frac{M}{r}
-r(lnP)_{u},
\ee
where \(\dot{D}=dD/du\). In the new gauge the field equations (4.11) and 
(4.12) become
\begin{eqnarray}\label{7.4}
\Delta \Delta (lnP)+12b^{2}M(lnP)_{u}-4b^{2}M_{u} 
+D^{2}[2((lnP)_{u})^{2}-(lnP)_{uu}]\nonumber\\ 
+D \ddot{D}-3D \dot{D}(lnP)_{u} 
= \frac{2b^{2}P^{2}}{MD}M_{z}M_{\bar{z}}~,
\end{eqnarray}
\be\label{7.5}
\Delta Q= D P^{2}(Q/P^{2})_{u}.
\ee
When \(D\) is taken to be constant, \(D=k\), these equations as well as (7.1)-
(7.3) reduce to their previous forms. Moreover, if \(D\) is any given function 
of \(u\), one can pass to the gauge of the Section IV by the coordinate 
transformation
\be\label{7.6}
\tilde{r}= \frac{k}{D(u)} r~,~~\tilde{u}= \frac{1}{k} \int D(u)du,
\ee
which has a unit Jacobian. If one also introduces
\be\label{7.7}
\tilde{H}= \frac{k^{2}}{D^{2}}H+ \frac{k \dot{D}}{D^{2}} \tilde{r},
\ee
\be\label{7.8}
\tilde{P}= \frac{k}{D}P~,~~\tilde{M}= \frac{k^{3}}{D^{3}}M~,~~\tilde{Q}= 
\frac{k^{2}}{D^{2}}Q
\ee
then it can be checked that (4.5)-(4.12) are all valid for the tilded 
variables.

Suppose now \(Q\) is a constant: \(Q=Q_{0}\). Then \(M=Q_{0}^{2}/D(u)\) 
can depend only on the null coordinate \(u\). Let us also assume that 
\(P=P(z,\bar{z})\) and specialize once again to the case where the Gaussian 
curvature \(K= \Delta lnP\) is constant. Under these assumptions,
\be\label{7.9}
\Phi_{1}=- \frac{i}{2 \tilde{\kappa}} \frac{Q_{0}}{r^{2}}e^{2\p}~,~~
\Phi_{2}=0~,
\ee
\be\label{7.10}
2H= \frac{1}{b}(K- \dot{D})- \frac{2M}{r}~,
\ee
\be\label{7.11}
ds^{2}=2dudr+2Hdu^{2}-r^{2}e^{-2\p}d \Omega_{2}^{2},
\ee
where \(d \Omega_{2}^{2}\) is the line element for the two-dimensional space 
of constant Gaussian curvature. The field equation (7.5) is then trivially 
satisfied and (7.4) reduces to an ordinary differential equation for \(D(u)\):
\be\label{7.12}
D^{3} \ddot{D}+4b^{2}Q_{0}^{2} \dot{D}=0.
\ee
Any solution of (7.12) will give us a particular member of the Robinson-
Trautman family and the simplest solution is, of course, \(D=k\). 
Choosing for convenience \(b=1\), one regains in this particular case the 
solutions (4.15) with a constant \(M\). The \(K=1\), \(\dot{D}=0\) solutions 
are known to possess both future and past null infinities  as 
well as a spatial infinity and describe black holes. Let us 
therefore concentrate on the \(K=1\), \(\dot{D} \neq 0\) solutions which are 
not gauge equivalent to the black hole solutions. It is easy 
to see that such solutions will be still asymptotically flat in the sense that 
they will possess at least a portion of the future null infinity. 
The parameter \(Q_{0}\) will be interpretable as the conserved magnetic  
charge of the solutions. Using the fact that the dilatonic current:
\be\label{7.13}
\star j_{D}~=~\star d \p- \tilde{\kappa}^{2}e^{-2\p}A \wedge \star F,
\ee
where \(A=A_{\m}dx^{\m}\), is conserved by the virtue of the field equation 
(2.3), one can also relate \(D(u_{0})\) to the dilatonic charge on a 
\(u=u_{0}\) hypersurface but, of course, \(D(u)\) is not conserved. The same 
applies to the Bondi mass \(M_{B}(u)\) of the solutions which we define as 
\be\label{7.14}
M_{B}~=~M- \frac{1}{4b^{2}}D \dot{D}.
\ee
This definition is motivated by the field equation (7.4) and agrees with the 
ADM mass of the black holes at \(\dot{D}=0\). If \(\dot{D} \neq 0\), (7.4) 
implies that
\be\label{7.15}
\dot{M}_{B}=-(\dot{D})^{2}/4b^{2}
\ee
and consequently, \(M_{B}(u)\) is always a decreasing function of \(u\). Since 
\(D(u)\) can either increase or decrease as a function of \(u\), this means 
that the Bondi mass can either be converted into the dilatonic charge through 
the gravitational radiation or both of these quantities can be radiated out to 
the future null infinity.

Integrating (7.12) once gives
\be\label{7.16}
\dot{D}-2b^{2}Q_{0}^{2}/D^{2}=C,
\ee
where \(C\) is a constant and depending on whether \(C=0\), \(C>0\) or \(C<0\), 
one gets three distinct solutions. The positivity of the Bondi mass requires 
\be\label{7.17}
CD^{2} \leq 2Q_{0}^{2}b^{2},
\ee
and this condition is respected for all values of \(u\) if \(C \leq 0\). 
We shall display these solutions after choosing \(b=1\). Consider first the 
choice \(C=0\). Then
\be\label{7.18}
D(u)= \left(c_{0}+6Q_{0}^{2}u \right)^{1/3},
\ee
where \(c_{0}\) is an arbitrary real constant and the Bondi mass can be 
written as \(M_{B}=M/2\) where 
\(M=Q_{0}^{2} \left(c_{0}+6Q_{0}^{2}u \right)^{-1/3}\). 
For the remaining \(C \neq 0\) cases, \(D(u)\) involves two real constants 
\(c_{0},c_{1}\) and is given as an implicit function of \(u\). Letting 
\(C=2c_{1}^{2}\), one finds that
\be\label{7.19}
D(u)- \frac{Q_{0}}{c_{1}} tan^{-1} \left[ \frac{c_{1}}{Q_{0}}D(u) \right]
=c_{0}+2c_{1}^{2}u,
\ee
whereas for \(C=-2c_{1}^{2}\),
\be\label{7.20}
D(u)+ \frac{Q_{0}}{2c_{1}} ln \left| \frac{D(u)-Q_{0}/c_{1}}{D(u)
+Q_{0}/c_{1}}\right|=c_{0}-2c_{1}^{2}u.
\ee
Each of these solutions is of Petrov type D. As \(r \rightarrow \infty\), the 
only non-zero component of the Weyl spinor \(\Psi_{2}\) and the curvature 
scalar \(\Lambda\) behave as
\be\label{7.21}
\Psi_{2}=-(M-D \dot{D}/6)r^{-3}+O(r^{-4}),
\ee
\be\label{7.22}
-(\Psi_{2}+2 \Lambda)=(M-D \dot{D}/4)r^{-3}+O(r^{-4}),
\ee
and (7.22) can be used to check that (7.14) agrees with the general definition  
of the Bondi mass given e.g. in \cite{19}.

Using (3.15)-(3.17) it is, of course, possible to represent these solutions in 
alternative gauges where \(M\) is constant but \(D\) and \(Q\) are functions 
of \(u\) or where \(D\) is constant but \(M\) and \(Q\) are functions of \(u\). 
There is also a gauge in which (7.19) and (7.20) can be written in a unified 
manner as an explicit function of \(u\). This occurs when one sets 
\(\hat{M}(\hat{u})= \lambda \hat{Q}(\hat{u})\), 
\(\hat{D}(\hat{u})= \lambda^{-1}
\hat{Q}(\hat{u})\) for some real constant \(\lambda\). In this gauge the 
case \(C=0\) corresponds to
\be\label{7.23}
\hat{Q}(\hat{u})=4 \lambda^{3} \hat{u}+ \hat{c}_{0},
\ee
where \(\hat{c}_{0}\) is constant and if \(C \neq 0\), one obtains
\be\label{7.24}
\hat{Q}(\hat{u})= \hat{c}_{0}e^{\hat{C} \hat{u}}
-4 \lambda^{3}/\hat{C},
\ee
where \(\hat{C}=2\lambda C/Q_{0}\). Since \(M\) and \(D\) scale differently  
under (7.6), it is manifest in this gauge that none of these solutions is 
gauge equivalent to the stringy black holes.

Among these solutions (7.20) is particularly interesting because, in this case 
the integration constants can be arranged in such a way that {\it the solution 
approaches to the stringy black hole solutions at late retarded times}. This 
property can be easily seen also in the gauge of (7.24) by setting 
\(M_{0}= \hat{M}(\hat{u}= \infty)\) and \(Q_{0}= \hat{Q}(\hat{u}
= \infty)\) which implies that \(\lambda=M_{0}/Q_{0}\) and \(\hat{C}
=-4M_{0}^{3}/Q_{0}^{4}\). Let \(M_{0}>0\) which will be consistent with the  
fact that \(M_{B}\to M_{0}\) as \(\hat{u}\to \infty\). Then (7.24) gives
\be\label{7.25}
\hat{M}(\hat{u})=M_{0}+(\hat{c}_{0}M_{0}/Q_{0})exp(-4M_{0}^{3}
\hat{u}/Q_{0}^{4}),
\ee
\be\label{7.26}
\hat{Q}(\hat{u})=(Q_{0}/M_{0}) \hat{M}(\hat{u}),
\ee
\be\label{7.27}
\hat{D}(\hat{u})=(Q_{0}^{2}/M_{0}^{2})\hat{M}(\hat{u}).
\ee
In this coordinate system the Einstein frame metric is
\be\label{7.28}
ds^{2}=2d\hat{u}d\hat{r}+2\hat{H}d\hat{u}^{2}-(Q_{0}/\hat{Q})\hat{r}^{2}
e^{-2\p}d\Omega_{2}^{2},
\ee
where \(2\hat{H}=F(\hat{u},\hat{r})+1-2M_{0}/\hat{r}\)  and
\be\label{7.29}
F(\hat{u},\hat{r})=\hat{Q}/Q_{0}-1-\hat{r}\dot{\hat{Q}}/\hat{Q}
-Q_{0}\dot{\hat{Q}}/2M_{0}-2(\hat{M}-M_{0})/\hat{r},
\ee
with \(\dot{\hat{Q}}=d\hat{Q}/d\hat{u}\). The other fields are given by
\be\label{7.30}
e^{-2\p}=1-\hat{D}(\hat{u})/\hat{r},~~~~~~~~
\Phi_{1}=-ie^{2\p}\hat{Q}(\hat{u})/2\tilde{\kappa}\hat{r}^{2},~~\Phi_{2}=0.
\ee
After introducing the Kruskal coordinates \(\hat{U},~\hat{V}\):
\be\label{7.31}
\hat{U}=-e^{-\hat{u}/4M_{0}}
\ee
\be\label{7.32}
\hat{V}=e^{\hat{v}/4M_{0}}
\ee
where \(\hat{v}=\hat{u}+2\hat{r}+4M_{0}ln(\hat{r}/2M_{0}-1)\), one finds that
\be\label{7.33}
ds^{2}=\frac{32M_{0}^{3}}{\hat{r}}e^{-\hat{r}/2M_{0}}d\hat{U}d\hat{V}
+16M_{0}^{2}e^{\hat{u}/2M_{0}}F(\hat{u},\hat{r})d\hat{U}^{2}
-(Q_{0}/\hat{Q})\hat{r}^{2}e^{-2\p}d\Omega_{2}^{2}.
\ee
Since \(F(\hat{u},\hat{r}) \sim e^{-4M_{0}^{3}\hat{u}/Q_{0}^{4}}\) as 
\(\hat{u} \to \infty\), the solution converges exponentially fast to a 
solution having \(F(\hat{u},\hat{r})=0\), if 
\(Q_{0}^{2} < 2\sqrt{2}M_{0}^{2}\). 
For the stringy black holes \(Q_{0}^{2} \leq 2M_{0}^{2}\) and this condition 
is always obeyed. Therefore, by choosing \(Q_{0}^{2} \leq 2M_{0}^{2}\), the 
solution can be joined smoothly to a black hole solution of mass \(M_{0}\) and 
magnetic charge \(Q_{0}\) along the null hypersurface \(\cal{H}^{+}\) which is 
defined by \(\hat{u}=\infty\). The Penrose diagram describing  the Kruskal 
extension of this solution for \(Q_{0}^{2} < 2M_{0}^{2}\) is shown in Fig.1. 
Notice that, in contrast to the generic behavior of the vacuum 
Robinson-Trautman solutions, the extension of the present solution through the 
horizon \(\cal{H}^{+}\) is infinitely differentiable.
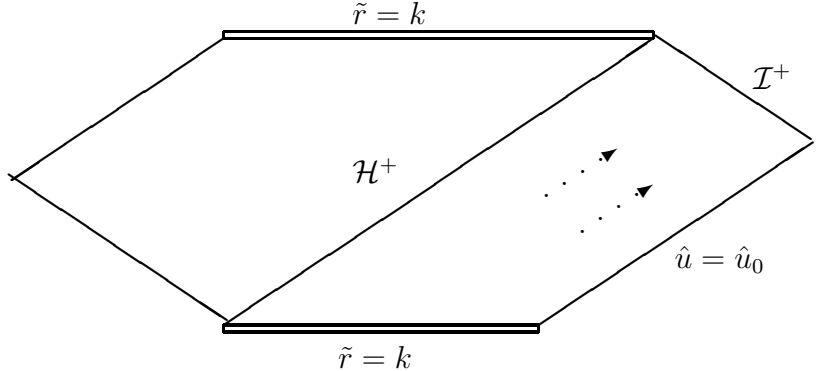
\begin{figure}{\centering{\setlength{\unitlength}{0.012500in}%
\begin{picture}(336,153)(30,612)
\thicklines
\put(120,750){\line( 1, 0){180}}
\put(300,750){\line( 0, 1){  3}}
\put(300,753){\line(-1, 0){180}}
\put(120,753){\line( 0,-1){  3}}
\put(120,630){\line( 1, 0){132}}
\put(252,630){\line( 0,-1){  3}}
\put(252,627){\line(-1, 0){132}}
\put(120,627){\line( 0, 1){  3}}
\put(120,630){\line( 3, 2){180}}
\put(120,750){\line(-3,-2){ 88.615}}
\put( 30,693){\line( 3,-2){ 91.385}}
\put(252,630){\line( 3, 2){114.923}}
\put(366,708){\line(-3, 2){ 65.077}}
\put(285,704){\vector( 3, 2){0}}
\multiput(255,684)(7.61538,5.07692){4}{\makebox(0.4444,0.6667){.}}
\put(300,689){\vector( 3, 2){0}}
\multiput(270,669)(7.61538,5.07692){4}{\makebox(0.4444,0.6667){.}}
\put(174,756){\makebox(0,0)[lb]{\smash{{$\tilde{r}=k$}}}}
\put(174,690){\makebox(0,0)[lb]{\smash{{$\cal{H}^{+}$}}}}
\put(168,612){\makebox(0,0)[lb]{\smash{{$\tilde{r}=k$}}}}
\put(309,654){\makebox(0,0)[lb]{\smash{{$\hat{u}=\hat{u}_{0}$}}}}
\put(342,729){\makebox(0,0)[lb]{\smash{{$\cal{I}^{+}$}}}}
\end{picture}
\label{fig1}
\caption{The Penrose diagram for the radiating solution which settles down to 
a stringy black hole when \(Q_{0}^{2} < 2M_{0}^{2}\). The singularities are 
located at \(\tilde{r}=k\), where \(\tilde{r}\) is the radial coordinate of 
Section 4. The \(\hat{u} \geq \hat{u}_{0}\) portion of the future null 
infinity is \(\cal{I}^{+}\) and \(\cal{H}^{+}\) is the future horizon.}
}}\end{figure}

\section{Conclusions}

\indent

In this paper we have studied a class of algebraically special solutions of 
the low energy string theory in four dimensions under the assumption that the 
spacetime admits a geodesic, shear-free, non-twisting null congruence. We have 
called all the solutions which follow from these assumptions the stringy 
Robinson-Trautman family and noted that the family possesses various 
remarkable properties. We have seen in particular that the family preserves 
its identity under the transformations between the string and the Einstein 
frames and the Lorentz Chern-Simons three-form was found to be always closed. 
In the detailed structure of the Robinson-Trautman family the crucial role was 
played by the dilaton field. This massless mode of the string theory regulated 
the divergence of the null geodesics and together with the spin-1 field gave 
rise to three subclasses of solutions. Two of these subclasses were seen to be 
diverging while the third one had a vanishing divergence. Due to this behavior
, it was possible to recover a significant number of the well-known solutions, 
including the stringy black holes and the pp-waves as particular members of 
the Robinson-Trautman family. We have also obtained some new solutions 
explicitly and described how a radiating solution tends smoothly to the 
stringy black holes at late retarded times.

Among the solutions we have studied the closest analogs of the vacuum
Robinson-Trautman solutions are the ones which belong to the first subclass
of Section IV. The field equations (4.10) and (4.11) which govern this
subclass reduce to the vacuum Robinson-Trautman equation when $k=Q=0$. The
vacuum Robinson-Trautman equation is known to be a special case of the
Calabi equation\cite{20} which arises in the study of the extremal K\"{a}hler
metrics and it will be interesting to see whether a geometric interpretation
can also be given for the stringy Robinson-Trautman equations (4.10) and
(4.11).

In Einstein's theory the global existence and the convergence of the solutions
of the vacuum Robinson-Trautman equation have been extensively studied
\cite{21}. These studies have shown that the vacuum Robinson-Trautman 
spacetimes exist for all positive retarded times and converge exponentially 
fast to the Schwarzschild manifold. The solutions, however, can be extended 
across the event horizon only with a finite degree of smoothness. How these 
properties are generalized in string theory and under what initial conditions 
the solutions converge to the stringy black holes is an interesting open 
problem. The appropriate framework for this problem is again furnished by the 
first subclass of Section IV. In this subclass a partial answer to the 
question of convergence of the solutions can be given easily by specializing 
to the case \(P=P(z,\bar{z})\). Under this assumption the parabolic 
equation (7.5) can be used to infer that all solutions converge to the 
\(Q=const.\) solutions. We have examined the \(Q=const.\) solutions whose 
Gaussian curvature is also constant in Section VII and found that there are 
three types of radiating solutions for the topology of a two-sphere. Only one 
of these solutions was seen to converge exponentially fast to the stringy 
black hole spacetimes and for this particular example the extension across the 
horizon was infinitely differentiable. Whether this is a generic behavior of 
all the solutions which belong to the first subclass of Section IV is not 
known at present and should be further investigated.

The radiating solution which converges to the \(Q_{0}^{2}=2M_{0}^{2}\) stringy 
black holes appears to be particularly interesting because both the electric 
\cite{7} and the magnetic\cite{22} extreme black holes can be interpreted as 
exact solutions of string theory and are supersymmetric\cite{23}. This raises 
the question whether the corresponding radiating solution is also exact and 
supersymmetric. Since a time-like Killing vector is not present, the radiating 
solution cannot be supersymmetric except at \(\hat{u}=\infty\). The issue of 
exactness, however, is more subtle and deserves further attention.

The principal goal of the present paper was to see how certain classical 
predictions of Einstein's theory are generalized in string theory. With all 
the above issues in hand, it seems reasonable to expect that stringy Robinson- 
Trautman family will furnish us with further insights towards the resolution 
of this fundamental problem.

\bigskip
\noindent
{\bf ACKNOWLEDGEMENTS}

\indent

The authors thank International Centre for Theoretical Physics, ICTP, Trieste,
for hospitality where certain portions of this work was done. The research 
reported in this paper has been supported in part by the Scientific and 
Technical Research  Council of Turkey, T{\"U}B{\.I}TAK and by the Turkish 
Academy of Sciences, T{\"U}BA.

\newpage

\appendix
\section{Appendix}
\indent 

In this Appendix we describe the NP variables in terms of exterior 
differential forms and compute the NP scalars for the stringy 
Robinson-Trautman metric (2.25). This formalism is similar to the 
approaches of \cite{24} and \cite{25} and was utilized in \cite{26}.

Let us adopt the NP spacetime conventions and represent the null tetrad 
in terms of a \( 2 \times 2 \) Hermitian matrix of one-forms 
\( \tilde{\sigma} \):
\be\label{A.1}
\tilde{\sigma}= \left( \begin{array}{cc} 
l & \bar{m} \\
m & n 
\end{array} \right).
\ee
The effect of a Lorentz transformation on  \( \tilde{\sigma} \) is

\be\label{A.2}                      
\tilde{\sigma} \to S \tilde{\sigma} S^{\dagger} ,
\ee
where \(S \in SL(2,C)\). Cartan's first equation of structure can be 
written as

\be\label{A.3}
d \tilde{\sigma}+ \tilde{\sigma} \wedge \Gamma 
- \Gamma^{\dagger} \wedge \tilde{\sigma}=0 ,
\ee
where

\be\label{A.4}
\Gamma= \left( \begin{array}{cc}
\Gamma_{0}  &  \Gamma_{2} \\
\Gamma_{1}  &  - \Gamma_{0}
\end{array} \right), 
\ee
is the  \(SL(2,C)\) Lie algebra valued connection. NP spin coefficients are 
defined as the coefficients of the complex connection one-forms:
\be\label{A.5}
\begin{array}{lll}
\Gamma_{0}&=& \gamma l+\epsilon n-\alpha m-\beta \bar{m},\\
\Gamma_{1}&=& -\tau l-\kappa n+\rho m+\sigma \bar{m},\\
\Gamma_{2}&=& \n l+\pi n-\lambda m-\m \bar{m}.\\
\end{array}
\ee
By writing (A.3) explicitly, it can be checked that

\begin{eqnarray}\label{A.6}
dl&=&-(\epsilon+ \bar{\epsilon})l \wedge n
+(\alpha+ \bar{\beta}- \bar{\tau})l \wedge m
+(\bar{\alpha}+ \beta- \tau)l \wedge \bar{m} \nonumber\\
  & &- \bar{\kappa}n \wedge m- \kappa n \wedge \bar{m}+(\rho- \bar{\rho})m 
  \wedge \bar{m},\nonumber\\
dn&=&-(\gamma+ \bar{\gamma})l \wedge n+ \n l \wedge m
+ \bar{\n} l \wedge \bar{m}+(\pi- \alpha- \bar{\beta})n \wedge m\nonumber\\
  & &+(\bar{\pi}- \bar{\alpha}- \beta)n \wedge \bar{m}
+(\m- \bar{\m})m \wedge \bar{m},\\
dm&=&-(\tau+ \bar{\pi})l \wedge n+(\bar{\m}+ \gamma- \bar{\gamma})l \wedge m
+ \bar{\lambda} l \wedge \bar{m}\nonumber\\
  & &- \sigma n \wedge \bar{m}
+(\epsilon- \bar{\epsilon}- \rho)n \wedge m
+(\beta- \bar{\alpha})m \wedge \bar{m},\nonumber
\end{eqnarray}

The \( SL(2,C)\) Lie algebra valued curvature is defined as

\be\label{A.7}
R=d \Gamma+ \Gamma \wedge \Gamma=\left( \begin{array}{cc}
\Theta_{0} & \Theta_{2} \\
\Theta_{1} & -\Theta_{0}
\end{array} \right),
\ee
where the curvature two forms

\begin{eqnarray}\label{A.8}
\Theta_{0}&=& d \Gamma_{0}- \Gamma_{1} \wedge \Gamma_{2}, \nonumber \\ 
\Theta_{1}&=& d \Gamma_{1}-2 \Gamma_{0} \wedge \Gamma_{1},\\
\Theta_{2}&=& d \Gamma_{2}+2 \Gamma_{0} \wedge \Gamma_{2} \nonumber
\end{eqnarray}
are expanded as

\begin{eqnarray}\label{A.9}
\Theta_{0}&=& (\Lambda- \Phi_{11}- \Psi_{2})l \wedge n+ \Psi_{3}l \wedge m
+ \Phi_{12}l \wedge \bar{m}- \Phi_{10}n \wedge m \nonumber \\
          & &- \Psi_{1}n \wedge \bar{m}
+( \Psi_{2}- \Phi_{11}- \Lambda)m \wedge \bar{m}, \nonumber \\
\Theta_{1}&=& (\Psi_{1}+ \Phi_{01})l \wedge n
- (\Psi_{2}+2 \Lambda)l \wedge m- \Phi_{02}l \wedge \bar{m}+ \Phi_{00}n 
\wedge m \nonumber \\
          & &+ \Psi_{0}n \wedge \bar{m}
+( \Phi_{01}- \Psi_{1})m \wedge \bar{m},\nonumber \\
\Theta_{2}&=&-(\Psi_{3}+ \Phi_{21})l \wedge n+ \Psi_{4}l \wedge m
+ \Phi_{22}l \wedge \bar{m} \nonumber \\
          & &- \Phi_{20}n \wedge m
- (\Psi_{2}+2 \Lambda)n \wedge \bar{m}
+( \Psi_{3}- \Phi_{21})m \wedge \bar{m}. 
\end{eqnarray}
Here \( \Psi_{0},\Psi_{1},\Psi_{2},\Psi_{3},\Psi_{4} \) are the components
of the Weyl spinor, \(24 \Lambda\) is the curvature scalar and \( \Phi\)'s  
represent the Ricci spinor \cite{15}.

Let us define the Hodge duals:
\be\label{A.10}
\begin{array}{lll}
\star 1=-il \wedge n \wedge m \wedge \bar{m}, &  & \\
\star l=il \wedge m \wedge \bar{m},& \star n=-in \wedge m \wedge \bar{m}, & 
\star m=-il \wedge n \wedge m, \\
\star (l \wedge n)=im \wedge \bar{m},& 
\star (m \wedge \bar{m})=il \wedge n,& \\
\star (l \wedge m)=-il \wedge m,& 
\star (n \wedge \bar{m})=-in \wedge \bar{m},&
\end{array}
\ee
and note that \(\star \star 1=-1,~~ \star \star l=l\). In these conventions
one finds that for a vacuum spacetime, \( R_{\m \n}=0 \):

\be\label{A.11}
\star R =-iR.
\ee

Consider now the metric (2.25). When the tetrad is choosen as in (2.25),
the non-\-ze\-ro spin co\-ef\-fi\-cients are
\be\label{A.12}
\begin{array}{ccc}
\gamma=H'/2, & \alpha=-P_{\bar{z}}/2R, & \beta=P_{z}/2R,\\
\rho=-R'/R, & \n=-PH_{\bar{z}}/R, & \m=-[HR'/R+(ln(P/R))_{u}].
\end{array}
\ee

The non-zero components of the Weyl spinor are found to be

\begin{eqnarray}\label{A.13}
\Psi_{2}&=&\frac{1}{2}H''- \frac{1}{3} \left( \frac{HR'}{R} \right)'
- \frac{1}{6R^{2}} \Delta lnP + \frac{1}{3} \left( \frac{R_{u}}{R} \right)', 
\nonumber\\
\Psi_{3}&=&\frac{P}{2R} \left[(lnP)_{u \bar{z}}-H'_{\bar{z}}
+ \frac{2R'}{R}H_{\bar{z}}\right],\nonumber\\
\Psi_{4}&=&\frac{1}{R^{2}}(P^{2}H_{\bar{z}})_{\bar{z}},  
\end{eqnarray}
where  \( \Delta =2P^{2} \partial_{z} \partial_{\bar{z}}\).
The non-zero projections of the Ricci tensor are

\begin{eqnarray}\label{A.14}
\Phi_{00}&=&-R''/R, \nonumber \\
\Phi_{12}&=&- \frac{P}{2R}[H'_{z}+(lnP)_{zu}], \nonumber \\
\Phi_{11}-3 \Lambda&=& \frac{1}{2} H''- \left( \frac{R_{u}}{R}\right)' 
+ \frac{1}{R}(HR')'+ \frac{R'}{R}[ln(P/R)]_{u}, \nonumber \\
\Phi_{11}+3 \Lambda&=& \frac{1}{2R^{2}} \Delta lnP
                      + \left(\frac{R_{u}}{R}\right)' \nonumber \\
                   & &-H \left(\frac{R'}{R} \right)^{2}- \frac{1}{R}(HR')'
                   - \frac{2R'}{R}[ln(P/R)]_{u}, \nonumber\\
\Phi_{22}&=& \frac{1}{2R^{2}} \Delta H+(H'-2HR'/R)[ln(P/R)]_{u}
            + \frac{R'}{R}H_{u}-H^{2} \frac{R''}{R} \nonumber\\
         & &+2H \left(\frac{R_{u}}{R} \right)' +[ln(P/R)]_{uu}
            -[(ln(P/R))_{u}]^{2}.
\end{eqnarray}

\end{document}